\documentclass[conference, a4paper]{IEEEtran}
\IEEEoverridecommandlockouts
\usepackage{cite}
\usepackage{amsmath,amssymb,amsfonts}
\usepackage{algorithmic}
\usepackage{graphicx}
\usepackage{textcomp}
\usepackage{xcolor}
\def\BibTeX{{\rm B\kern-.05em{\sc i\kern-.025em b}\kern-.08em
    T\kern-.1667em\lower.7ex\hbox{E}\kern-.125emX}}
\usepackage{balance}

\begin{document}

\title{On licenses for [Open] Hardware\\
\thanks{978-1-7281-9132-4/20/\$31.00 \textcopyright 2020 IEEE}
}

\author{
    \IEEEauthorblockN{Màrius Montón\IEEEauthorrefmark{1}\IEEEauthorrefmark{2}, Xavier Salazar\IEEEauthorrefmark{3} }

    \IEEEauthorblockA{\IEEEauthorrefmark{1}Dpt. of Microelectronics and Electronics Systems, Universitat Autònoma de Barcelona, Spain.\\
    Email: marius.monton@uab.cat }
    
    \IEEEauthorblockA{\IEEEauthorrefmark{2}Institut d'Estudis Espacials de Catalunya. Barcelona, Spain}
    
    \IEEEauthorblockA{\IEEEauthorrefmark{3}Barcelona Supercomputing Center. Barcelona, Spain}
    
    % \IEEEauthorblockA{\IEEEauthorrefmark{4} 
    % Instituto de Microelectrónica de Barcelona, IBM-CNM (CSIC), Spain}
}

\maketitle

\begin{abstract}
This document explains the basic concepts related to software and hardware licenses, and it summarizes the most popular licenses that are currently used for hardware projects. Two case studies of hardware projects at different levels of abstraction are also presented, together with a discussion of license applicability, commercial issues, code protection, and related concerns.

This paper intends to help the reader understand how to release open hardware with the most appropriate license, and to answer questions that are of current interest. We have been mainly motivated by the growing influence of the open RISC-V ISA, but trying to address a wider hardware point of view.
\end{abstract}


%%%%%%%%%%%%%%%%%%%%%%%%%%%%%%%%%%%%%%%%%%%%%%%%%%%%%%%%%%%%%%
\section{Introduction}
\label{sec:intro}

Thanks to the success and widespread adoption of many open-source projects (Linux, LibreOffice, etc.), the community and users typically have a good understanding of software licenses. Developers in general have a clear picture of the overall landscape of software licenses, together with the pros and cons and rights retained and conceded by each type of software license.

On the other hand, when we switch to hardware projects, the landscape of licenses while being generally and conceptually similar they can vary lightly and the knowledge of developers is scarce \cite{ackerman2008toward, greenbaum2011open, katz2019survey, svorc2019breathe, luis2019open}. With the advent with the advent of of open source Hardware initiatives \cite{riscv-isa} it is getting more important to become familiar with those concepts. Hence, this paper tries to address this challenge by clarifying the basic concepts of licensing, how they are used for hardware and it applies this knowledge to two different use cases \cite{asanovic2014instruction, 10.1145/3386377}. 
%add cite -> https://dl.acm.org/doi/10.1145/3386377 
%add cite -> https://people.eecs.berkeley.edu/~krste/papers/EECS-2014-146.pdf
%MMM done

The paper is structured as follows: Section~\ref{sec:basics} introduces a number of key basic concepts related to licenses. Section~\ref{sec:licensetypes} is a shortlisting of the most commonly used licenses at the moment, with their main features. Next, Section~\ref{sec:casestudy} explores the two use cases. The article finishes with the conclusions in Section~\ref{sec:conclusions}. 

%%%%%%%%%%%%%%%%%%%%%%%%%%%%%%%%%%%%%%%%%%%%%%%%%%%%%%%%%%%%%%
\section{License basics}
\label{sec:basics}

In order to understand open hardware licensing it is important to know main basic concepts of open source licensing that have been traditionally used for software as many of those concepts can be later on extrapolated for hardware. 
Licenses are the legal contract between the Software owner and the users. It usually specifies the permissions granted to the user (\textit{the licensee}) and the rights, obligations and limitations of the user towards the Software. Typically a software license grants the licensee to use the software but not to re-distribute or re-sell the software.
The level of rights retained by the owner of the software classifies what type of license to use. For instance, Microsoft Windows has a proprietary license that retains copyright to Microsoft, but gives permission to the licensee (end-user) to use the software and show the software to others. But the same license doesn't allow users to copy the software (install it on other computers), modify the software or re-distribute to others. GNU/Linux, has a license (GPL - General Public License) \cite{GPL} that allows the end-user to access the code, copy the software, modify and re-distribute it (with some restrictions).

First of all, some definitions of common concepts used in licenses are described below, emphasizing the differences between terms that are often misused: 

\subsection{Free and Open}
\label{sub:FreevsOpen}
As mentioned before, there are two concepts that are often misunderstood: free and open. \textit{Free} means that the licensee gets the Software free-of-charge (i.e. "gratis"), also known as libre software, and \textit{Open} means that the user gets access to the source code of the design. Although these two concepts are quite different, they are usually seen together.\footnote{There are examples of Software that is free but not open (i.e. Winrar) and software that is open but not free (i.e. $\mu$C/OS-III)}

\subsection{Proprietary and Open}
\label{sub:ProprietaryvsOpen}
 In contrast to free and open approaches, the code may be kept proprietary, closed as an industrial or trade secret. In these cases, only software binaries or encrypted code are shared. The use of proprietary software is often regulated by a limited use license that may include non-disclosure clauses, prevent reverse engineering, or add any further protection mechanisms such as liability clauses. Alternatively software or the underlying methodology may also be patent protected, and therefore disclosed. Both cases can become relevant when assessing their compatibility with open source components 

\subsection{Authorship and Ownership}
\label{sub:AuthorshipvsOwnership}
The terms authorship and ownership are also often mixed and misunderstood. The authors are the software development team that wrote the code. The owner is typically the institution that employed the authors. 

\subsection{Software disclosure}
Each organization has its own procedure for software disclosure. The authors are not necessarily responsible for choosing the license of the codes, but they must follow the rules of their organizations depending on the business/academic strategy. Specific terms are usually regulated in the employees' working or service contracts.

\subsection{Derivative work}
% https://www.linuxjournal.com/article/6366
A derivative work is any work that expands or uses previous work from another author. 
In the case of software, it is commonly accepted that any code that uses a library by statically linking it is a derivative work of that library. 
An exception to this definition is if a Software component used is designed to be a library and it is licensed "as is".
In the case of hardware, a design that uses a module or parts of a module can be considered as derivative work of this module.

\subsection{Copyright, Creative Commons and Public Domain}
All software is automatically protected by copyright to its author, giving exclusive rights to determine under what terms the original work may be copied and used by others.

An author can give away ("waive") its copyright rights by offering these rights to the public domain, giving freedom to people to distribute copies, modify it, or sell it without attribution.

Creative Commons is a set of licenses intended for the distribution of documents or content of any type in a copyleft philosophy \cite{cc}. There are variations of the rights left to the licensee, but basically are all combinations of the following:
\begin{itemize}
    \item CCO means "no right reserved" enabling waiving the copyright. 
    \item Attribution: Licensee can distribute, modify, remix, sell, etc. but they must always give  credit to the original author.
    \item ShareAlike: Same as above, but the license of the derivative work must use the same license.
    \item NoDerivs: Limit the licensee to distributing derivative work.
    \item NonCommercial: The licensee cannot use the work commercially.
\end{itemize}

\subsection{Copyleft/Reciprocal and Permissive}

Copyleft is the practice of offering the rights to freely distribute copies or modified versions of a work with the condition that the same rights must be preserved in the derivative works. These are often called viral licenses.

In contrast, permissive licenses allow the licensees to redistribute  derivative works with greater freedom, without demanding that the licensee publish derivative works with the same license.

In the case of hardware, as it is not covered by copyright, it is preferable to use the word reciprocal to express the same Copyleft concept.

\subsection{License compatibility}
License compatibility is the capacity of pieces of software or hardware (IP blocks) with different licenses to be distributed together. 

Two incompatible licences can contain contradictory statements and it can be impossible to legally combine source code from different software to create or publish a new software.

\subsection{Patents and open-source}

Integrating patent-protected software with free and open-source software is a special case of compatibility. Free software or open source projects cannot fulfill patent licences that include any kind of per-copy fee. A patent that is royalty-free is acceptable.

%%%%%%%%%%%%%%%%%%%%%%%%%%%%%%%%%%%%%%%%%%%%%%%%%%%%%%%%%%%%%%
\section{License types}
\label{sec:licensetypes}

In this section license needs and types are introduced.

\subsection{Philosophy}

\begin{figure}
  \centering
  \includegraphics[width=\linewidth]{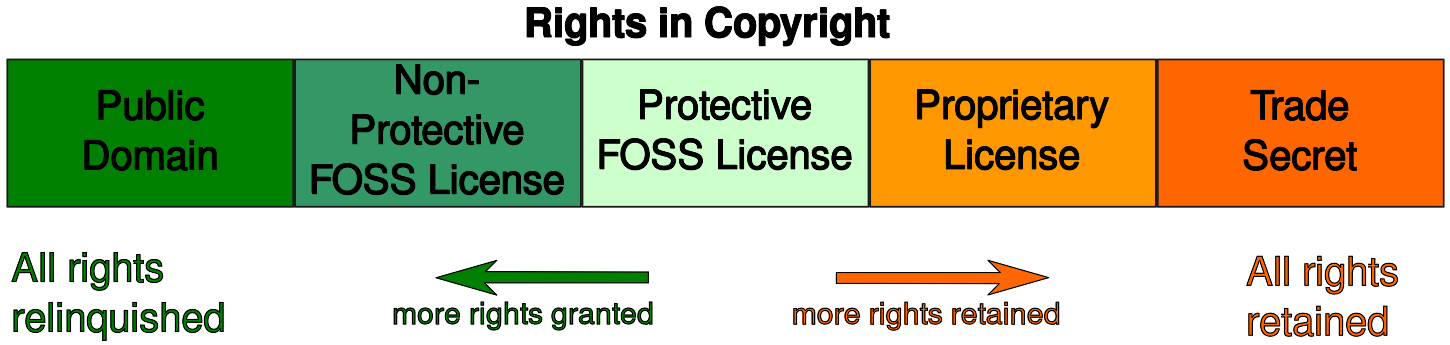}
  \caption{Degrees of freedom and rights in SW licenses. \cite{licenses_diag}}
  \label{fig:licenses}
\end{figure}

Proprietary licenses try to protect a business and a value created by the company or companies and their developers. For this reason, this type of licenses maintains the copyright and controls in some way the ability of the user to get additional copies or  re-distribute the Software. Basically, the Software sold has a value, and the user must pay for each piece or installation. The commercial licenses try to protect this selling strategy.
The majority of code of this type is sold as "closed-source", that is, the customer doesn't receive a copy of the source code of the Software bought, but only the binary files that are necessary to run the software.
In this way, the selling company has double protection, because the license doesn't allow the customer to modify or copy the software and technically the customer without the source code has a  limited capacity to do so.\footnote{There are companies that sell Software in source code form using proprietary licenses.}

The free and open-source software philosophy (also named FOSS) tends to grant the user the ability to inspect, modify and distribute the Software. Different degrees of freedom are given to the end-user by different FOSS licenses, as described in Section~\ref{sub:SWlicenses}. A depiction of this concept is shown in Figure~\ref{fig:licenses}.

Although free and open licenses give  the end-users more freedom than commercial licenses, some of them restrict in some way the options of the end-users to do whatever they want to the code. For instance, GPL restricts the licensee to use the same type of license in any derived works. 

In terms of commercialization, Open Source codes are in most cases given ``as is'', without warranties, support or maintenance and they are not business friendly. All responsibility of the correct use of the Software relies solely on the user of the licence. It is therefore possible to retain business value from open source technologies by offering services around them instead of simply selling the resulting software product. Hybrid approaches are also possible, with the choice of dual licences that include open source sharing for limited (non-commercial) purposes, but with a proprietary approach for commercial uses.

\subsection{Software licenses}
\label{sub:SWlicenses}

As said, there are several different licenses designed for the distribution of software. Below the most common ones are presented.

\subsubsection{GPL}
\label{sub:GPLlicense}
General Public License (GPL) refers to the licenses designed and used by the GNU project \cite{GPL}. It is a free license that gives the end-user the freedom to run, study, share and modify the software. 

This license is copyleft, meaning that the derivative work can be only licensed with the same GPL license.

It is the license used by Linux Kernel and the GCC compiler and it is one the most popular software license in FOSS domain.

There exists a variant of the GPL license intended for libraries, which is called LGPL (GNU Lesser General Public License). This license allows non-GPL code to be linked with libraries licensed by LGPL. 

\subsubsection{BSD licenses}
\label{sub:BSDlicense}
% https://www.freebsd.org/doc/en_US.ISO8859-1/articles/bsdl-gpl/article.html#bsd-advantages
% Comparing the BSD and GPL Licenses https://timreview.ca/article/67
BSD (Berkeley Software Distribution) licenses are a set of FOSS licenses less restrictive than GPL \cite{BSD}. There are multiple forms of this license, but the most common in use is the license known as ``3-clause version''.
These licenses allow the licensee to use, run, study, share and modify the software. 

What this license doesn't do is to force the publication of derivative work as FOSS, but it requires that credit is given to the original authors. It means that derivative works can be closed and commercialized in a ``standard way''.

\subsubsection{MIT}
\label{sub:MITlicense}
% https://mit-license.org/
The MIT (Massachusetts Institute of Technology) license is very similar to BSD license with the same level of permission granted to the licensee \cite{MIT}.

It is compatible with other free or copyleft licenses and allows to reuse within proprietary software.

\subsubsection{EUPL}
\label{sub:EUPLlicense}
% https://ec.europa.eu/info/european-union-public-licence_en
% https://joinup.ec.europa.eu/collection/eupl/matrix-eupl-compatible-open-source-licences
European Union Public Licence (EUPL) is a new license created by the European Commission for FOSS~\cite{eupl}. 
This license is a copyleft type, allowing licensee to modify, redistribute, do derivative work, etc. 

It is compatible with GPL. 

\subsubsection{Apache License}
\label{sub:Apachelicense}
% http://www.apache.org/licenses/
The Apache License by the Apache Software Foundation (AFL) is a permissive free open-source license \cite{apache}.

% analysis to explain the summary table
% check if reference can be found

\subsection{Summary table}

Table~\ref{tab:SWsummary} In all of them regulation of rights to copy, distribute, modify, distribute a derivative and protection against patent claims summarizes what a user can do and don't for each license introduced in this document.

\begin{table}
\caption{License summary}
\begin{center}

\begin{tabular}{|c|c|c|c|c|c|}
\hline
\textbf{License} & \textbf{GPL} & \textbf{BSD} & \textbf{MIT} & \textbf{EUPL} & \textbf{Apache} \\
\hline
Copyright & Yes & Yes & Yes & Yes & Yes \\
Distribute & Yes & Yes & Yes & Yes & Yes\\
Modify & Yes & Yes & Yes & Yes & Yes\\
Derivative & Yes & No & No & No & No \\
Derivative license & GPL & Any & Any & Copyleft & Any \\
Patent grant & Yes & No & No & Yes & Yes\\
\hline
\end{tabular}
\label{tab:SWsummary}
\end{center}

\vspace{0.3 cm}
\begin{itemize}
 \item Copyright: Copyright is retained by original authors
 \item Distribute: Licensee can distribute the source code among others.
 \item Modify: Licensee can modify the source code
 \item Derivative: Derivative work must be shared in source code
 \item Derivative license: Derivative work license type
 \item Patent grant: protect both sides from patent claims
\end{itemize}
\end{table}

\subsection{Compatible licenses with other licenses}
\begin{figure}
  \centering
  \includegraphics[width=\linewidth]{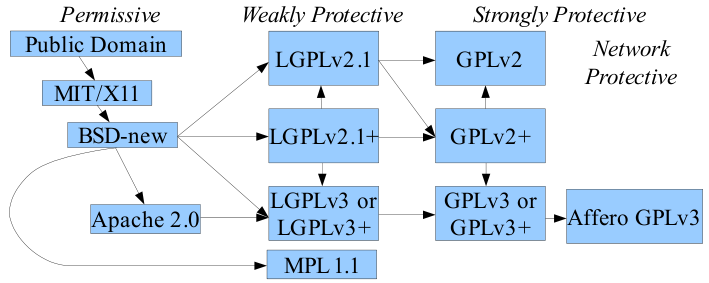}
  \caption{License compatibility diagram. (David A. Wheeler \cite{compatibility} )}
  \label{fig:compatibility}
\end{figure}

Figure~\ref{fig:compatibility} shows a diagram with compatibility relations between licenses. An arrow from A to B means that modules using licences A and B can be combined and the result has to use the license B.

This diagram is useful when using 3rd party modules or sub-projects with a different licenses than that chosen by the overall project. Hence, if we are licensing our code with a BSD-like license and we are using a module or library that is licensed with an LGPL license, any resulting project that integrates both parts must be licensed with LGPL (there is an arrow from BSD-new to LGPL in  Figure~\ref{fig:compatibility}).

\subsection{Software associations}

\subsubsection{Free Software Foundation}
\label{sub:FSF}
The Free Software Foundation (FSF) is a nonprofit with a worldwide mission to promote computer user freedom \cite{fsf}. 

\subsubsection{The Open Source Initiative}
\label{sub:OSI}
The open source initiative (OSI) aims to raise awareness and adoption of open source software, and build bridges between open source communities of practice \cite{osi}.

\subsection{Hardware licenses}
% https://en.wikipedia.org/wiki/Open-source_hardware
% https://fossi-foundation.org/
% Eli Greenbaum (2011) Open Source Semiconductor Core Licensing - Harvard Journal of Law & Technology Volume 25, Number 1 Fall 2011: http://jolt.law.harvard.edu/articles/pdf/v25/25HarvJLTech131.pdf

Hardware can be open-sourced in a similar way to software.% As in software, we named open-source the feature to publish and make available the source code of a software, hardware can also be open-sourced. 
In the case of hardware, the term ``open-source'' refers to the availability of all files and components necessaries to replicate the device. 

The basic hardware unit subject to be licensed is called an IP block (Intellectual Property block) or IP core. It is the reusable unit of logic, cell, netlist or integrated circuit layout design. The IP block also includes all the documentation necessary to understand its design or to use the block.

In the case of a PCB board, an open-source hardware release will contain the schematic and the layout in a free or standard format. Hardware files are usually also released in some format readable to all as PDF. For this different nature of the devices, there exists a set of licenses designed for hardware. Some projects and associations recommend general open-source licenses that were originally conceived for distribution of software, such as GPL, BSD or MIT.\footnote{``electrons are cheap, but atoms are expensive''} Usually these licenses only protect the hardware part (named the Documentation), and allow the author to chose what license to use for the software parts (Firmware, other software, etc.).

In the case of FPGA-based design, the release would include all HDL files (Verilog, VHDL, SystemC, etc.) in text format, etc. Although these kind of files can be licensed using software licenses, the use of hardware licenses is recommended as the terminology in hardware licenses fits better for this purpose. 

For a low-level design (Hard-IP, ASIC designs, FPGA bitstreams, etc.), software licenses are not enough and hardware licenses ought be used because of the nature of the files and the design process itself.

The Open Source Hardware Association (OSHWA) has a definition of Open Hardware in its webpage.\footnote{https://www.oshwa.org/definition/} See also Section~\ref{sub:OSHA}.

Below is a short listing of the commonly used open source hardware licenses:

\subsubsection{CERN Open Hardware License}
% https://ohwr.org/project/cernohl/wikis/home
% https://cern-ohl.web.cern.ch/
The CERN Open Hardware License (OHL or CERN OHL) was originally written by CERN to publish some of their designs \cite{cernohl}. There are three regimes for this license:
\begin{itemize}
    \item CERN-OHL-S: strongly reciprocal license. It means that a licensee can use the original code, and any releases of binary files (e.g. FPGA bitstreams incorporating the original Verilog code) must also provide all the HDL code and necessary components as well with the same license. 
    \item CERN-OHL-W: weakly reciprocal license. Similar to CERN-OHL-S, except that the original project can be published without including third-party components (but these components must be available to anybody, maybe by purchasing them). 
    \item CERN-OHL-P: permissive license. This licence gives total freedom to the licensee, it can be re-licensed without redistributing the original code.
\end{itemize}

\subsubsection{Solderpad}
The Solderpad Hardware Licence is based on the Apache 2.0 software license, adjusted to the hardware context. It is a permissive free open-source license \cite{solderpad}.

\subsection{Hardware associations}

\subsubsection{Open Source Hardware Association}
\label{sub:OSHA}
\begin{figure}
  \centering
  \includegraphics[width=0.2\linewidth]{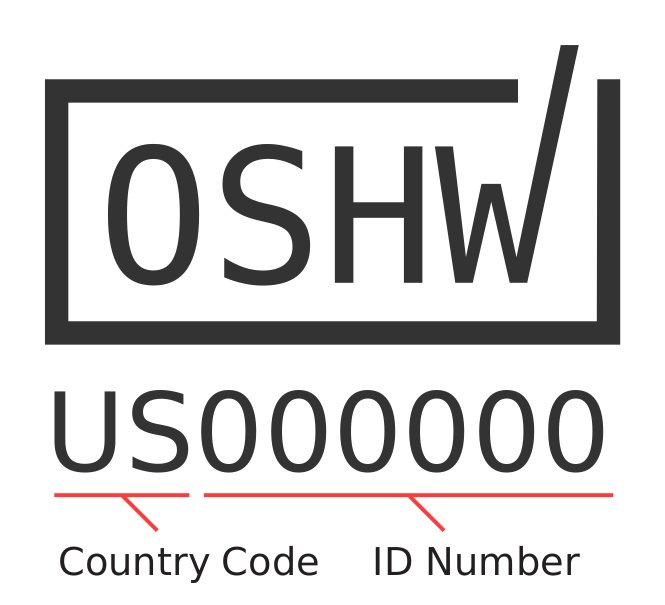}
  \caption{Open Source HW Association Certification logo}
  \label{fig:oshwa}
\end{figure}

The Open Source Hardware Association (OSHA) promotes access to all kinds of hardware and projects to users. It has a certificate program and maintains a database of open source hardware projects~\cite{OSHA, OSHAcert}. 
Each certificated project has a unique identifier and the project components (hardware PCBs, manuals, etc.) can show the OSHW logo (Figure~\ref{fig:oshwa}).

\subsubsection{FOSSi foundation}
The Free and Open Source Silicon Foundation (FOSSi) is a organization built to promote and assist free and open digital hardware designs\cite{fossi}. It maintains a repository of free hardware IP cores and promotes some projects. 

Table~\ref{tab:HWsummary} summarizes in the same way as previous table the hardware licenses.

\begin{table}
\caption{Hardware Licenses summary}
\begin{center}

\begin{tabular}{|c|c|c|c|c|c|}
\hline
\textbf{License} & \textbf{CC0} & \textbf{OHL-S} & \textbf{OHL-W} & \textbf{OHL-P} & \textbf{Solderpad} \\
\hline
Copyright & No & Yes & Yes & Yes & Yes \\
Distribute & Yes & Yes & Yes & Yes & Yes\\
Modify & Yes & Yes & Yes & Yes & Yes\\
Derivative & No & Yes & Yes & No & No \\
Deriv. license & Any & OHL-S & OHL-W & Any & Any \\
Patent grant & No & Yes & Yes & Yes & Yes\\
\hline
\end{tabular}
\label{tab:HWsummary}
\end{center}

\vspace{0.3 cm}
\begin{itemize}
 \item Copyright: Copyright is retained by original authors
 \item Distribute: Licensee can distribute the source code among others.
 \item Modify: Licensee can modify the source code
 \item Derivative: Derivative work must be shared in source code
 \item Derivative license: Derivative work license type
 \item Patent grant: protect both sides from patent claims
\end{itemize}
\end{table}

% https://www.oshwa.org/definition/
% https://certification.oshwa.org/
% https://www.oshwa.org/sharing-best-practices/

\subsection{Licensing the documentation}

Documentation, as a literature work, is different in nature to software or hardware, so a different type of license is required. Here are presented the most used licenses for documentation.

\subsubsection{GFDL}
GNU Free Documentation License is a copyleft license for documentation \cite{GFDL}. Its terms are similar to GPL, because it gives similar rights to the readers to copy, redistribute and modify a work, and the derivative work must have the same license. The original author can mark some parts of the text as invariants and the license forbids the licensees to modify theses parts.

Wikipedia uses this license for its articles.

\subsubsection{Creative Commons}
\label{sub:cc}
Anther option is to use a Creative Commons license. Some Open Hardware projects use ``Creative Commons Attribution Share-Alike'' to license their design files (schematics and layout files) (see Section~\ref{sub:cc}). It allows the licensee to sell derivative works.\footnote{Arduino and other companion business use this license for hardware files}

% HW deliverables / HW Outcomes subject to be IP protected

% \subsection{Practical issues}
% % Using the GPL/LGPL for HDL designs Open questions from a designer's perspective Javier Serrano August 1, 2014 https://ohwr.org/project/ohr-meta/uploads/5f52fc9362c7a8056ad92ac7828d2122/gpl_for_hdl.pdf 
% In this section we summarize some practical aspects of using licenses, far beyond the legal and formal aspects.

% 1.- The use of a license is mandatory in case the project is released in "open-source" form. 

% 2.- Every file of the project must have a header with copyright notice and license mention. In the repository a file with the entire license text must be present. This file is usually named COPYING or LICENSE.

% 3.- 

% \subsection{FAQ}

% Q1:

% A1: 

% Q2: Can ``switch'' from a open license to a closed license once the project is released in open form?

% A2: Yes, because author keeps copyright, he can change license at any time.

% Q3: A customer that uses our GPL SW and do some local changes, must publish its modifications?

% A3: No, if the customer doesn't sell/distribute the modified version of our code.

\subsection{Hardware deliverables of IP protectable assets}
The deliverables of a typical hardware project can be: the layout ready to tape-out to a ASIC vendor (e.g. TSMC, XFAB, or GF); the P \&R (place and route) project ready to be imported into a ASIC vendor tool (e.g. the GDSII file); the schematic or netlist in some vendor tool format (i.e. Cadence, Mentor). The outcome of a hardware project can also be a set of Gerber files for a PCB layout.
For an RTL project, Verilog or VHDL files are the foundation of the project (although higher level languages as SystemVerilog, SystemC or Chisel are getting more importance), but intermediate files as netlist (EDIF format) or final files as bitstreams can also be the outcomes of the project.

% Can we make a table for the analogies:

\subsection{Analogy from software to hardware licenses}
% Copyleft: GPL alike - CERN-OHL-H
% Lesser copyleft: LGPL - CERN-OHL-W
% permisive: MIT / APACHE / BSC - CERN-OHL-P / Solderpad
% IP block - library / code
% Synthesis - Compile
% bitstream - binary
% 
Regarding compatibility, the chart for open source Software can be extrapolated to the Hardware ones as summarized in Table~\ref{tab:SWHWsummary}. %\paul{Does this refer to Figure~\ref{fig:compatibility}? If there is a picture like Figure~\ref{fig:compatibility} for the hardware licences it would be great!!}

\begin{table}
\caption{SW / HW summary}
\begin{center}
\begin{tabular}{|c|c|}
\hline
\textbf{SW} & \textbf{HW} \\
\hline
Copyleft GPL alike & CERN-OHL-H \\
Lesser copyleft: LGPL & CERN-OHL-W \\
Permisive: MIT / APACHE / BSC & CERN-OHL-P / Solderpad \\
Library / code & IP block\\
Compile & Synthesis \\
Binary & bitstream \\

\hline
\end{tabular}
\label{tab:SWHWsummary}
\end{center}
\end{table}

%%%%%%%%%%%%%%%%%%%%%%%%%%%%%%%%%%%%%%%%%%%%%%%%%%%%%%%%%%%%%
\section{Case study examples}
\label{sec:casestudy}

In this section we propose two different hardware project approaches and study some of the implications related to the choice of license. The two projects are at different abstraction levels (the first is RTL only, and the second mixes RTL with gates and ASIC tools). Both projects are speculative and with the purpose of discussing hypotheses around the different licensing choices and their consequences

\subsection{Microprocessor at RTL level}
% needs some rewording - authors not necessarily are the ones that decide on the licence to set (it depends in the institution they are working). 
This case study is for a RISC-V microprocessor in Verilog language at RTL (Register-Transfer Level) \cite{riscv-isa}. The code is intended to be synthesized to FPGA and it uses some vendor-library or primitives, such as embedded memories, and DSP blocks in FPGA.
Also, the code has been synthesized to an ASIC to test correctness and  suitability of the written code to the technology.  
The development team as authors hold the copyright of the entire code, but the ownership is with their institution, and therefore any decision on IP disclosure has to be consented by institution's decision makers. 
There are many commercialization and licensing options, which we enumerate, from those allowing more freedom to those that are more restrictive.

\subsubsection{Public domain}
%They can chose% 
It can be considered to publish all Verilog files and auxiliary files (simulation scripts, FPGA synthesis scripts for one specific vendor, etc.) as copyleft with a Creative Commons CC0 license.

Once published with this kind of license, the code is no longer their property, and  it is now public. This means that any other team or developer can get the code or parts of it and use it without restrictions: sell it, use it in a product, etc. A company could get the code, do some changes to prepare an ASIC and sell it with any mention to the original authors.

\subsubsection{Reciprocal}
% https://ohwr.org/project/cernohl/wikis/FAQ#q-why-should-i-use-a-licence

%Original team decides to release 
The code can also be released into a public repository using the CERN-OHL-W license. In this case, any development team will be able to obtain the code, synthesize it and obtain the  same processor as the original, assuming that they have acquired the same vendor tools, primitives and libraries.  
The original team may not be able to use CERN-OHL-S if they use third-party libraries, if they cannot distributed with their project. If the entire project was designed using open-source with a license compatible with CERN-OHL-S, the entire project could be licensed with this license.

This licensee team can modify the original code and publish it again, but it will be mandatory to use the same license in the changes and/or new files as a derivatives. This team or organization can also release and sell the bitstream or a PCB board with the bitstream programmed in the FPGA and also has to provide the original code and his changes. If only parts of the design are taken and used in a design by another team, at least the copyleft parts must keep the same license.

\subsubsection{Permissive license}
%The team could chose 
A permissive license, such as the SolderPad license, can also be chosen. Using this license, the licensee team must publish the original code, but the licensee can choose not to publish its changes. While is it more difficult to get commercial value using this type of license, it is the most business friendly solution since the licensee has a higher level of freedom to use, modify and integrate in their systems without compromising the rights of the rest of the components.

\subsubsection{Non-commercial license}
It is also possible to publish the project with a license that forbids commercial use, but allows licensee to use, modify, etc. the original project for other purposes (mainly academia). This kind of license is very restrictive, because can be seen as ``nobody in a company can open the code'' and limits the scope of the project to just academia.

\subsubsection{Proprietary license}
%The original team 
It can also be decided to publish only the bitstream and release some documentation and publicity of its design, while keeping the source code unreleased. This would be a classical product with any public release of the code.

\subsection{Growing complexity of microprocessors}
Microprocessors are becoming more complex systems-on-chip, including heterogeneous architectures with general purpose processing units, specialized types of accelerators, memory systems, etc. In such cases, as many different IP blocks can be identified as part of a microprocessor, a more careful look at the compatibility of licenses is needed and therefore a more similar approach as the one explained in next section.

%%%%%%%%%%%%%%%%%%%%%%%%%%%%%%%%
\subsection{IP Block for ASIC}
This case study is for an analog--digital mixed design ASIC. There is a analog part that is designed with classical tools (schematic capture, full-custom layout and P\&R, etc.) but some of parts of the design are designed and parametrized using iterative processes using a high-level language as python. 
The IP has a add-on that is a digital interface to a processor bus (AXI4) written in VHDL without any third party libraries.
In this case the development team (original team) that holds the copyright of the entire code as authors, are also the owners of it.
Again, there are a number of licensing and publishing options, and we briefly discuss each.

\subsubsection{Public domain}
The original team can publish all files and license them with a Creative Commons CC0 license. 

\subsubsection{Reciprocal}
In case the original authors want to publish their work, it seems reasonable that they publish all design files, including schematics, P\&R results and other files involved in the ASIC synthesis. In this case, since ASIC libraries and toolkits are restricted to many people, the type of license is similar that the other study case. In this particular case, if using CERN-OHL, the license variant should be CERN-OHL-W, because the ASIC vendor toolkit and standard cell libraries are closed components.

For the VHDL code of the bus controller, it can be released with CERN-OHL-S because the original project is not using any third-party library, hence all code is available for licensees.

\subsubsection{Permissive license}
If the developers want to allow more flexible licensing for derivative works, they can choose a permissive license. 

In this case, all ASIC-related files can be published using CERN-OHL-W (for the same reason as the HDL case and the use of proprietary libraries or components) or using SolderPad license. 

For the digital RTL--VHDL part, a SolderPad license is the most suitable.

This is usually a good choice for handling complex schemes as it doesn't compromise  compatibility between components. On the other hand, it is possible that the original team  easily loses control of their original development.

\subsubsection{Non-commercial license}
Again, it is possible to publish the project with a non-commercial license, although is not recommended for the reasons exposed in the prior case. %\paul{Add cross-reference}

\subsubsection{Proprietary license}
The original team can chose to sell the IP has a hard-IP and the post-P\&R files as deliverables. In this case, probably the deliverable to protect will be the last steps of the ASIC design specific to a vendor-tool and a fabric technology  because the tools and design-kits are under restrictive NDAs for its exclusive use for academic purposes.

\section{Conclusions}
\label{sec:conclusions}
% some rewording 
In this paper we presented the different license types for software and hardware. A introduction to each of the licenses and a brief explanation and comparison for all of them has been exposed. Also, two case studies are explained. These two use cases work at different levels of abstraction and with different development tools. For each case different license options are depicted to show the landscape of possibilities and to help engineers to chose the proper license for their projects.

When deciding to publish a design following an open-source philosophy, it is important to realize that the main point should be to give others the ability to study, re-use or modify (and even sell) the project outcomes. For that reason, it is important to publish not only the intermediate or final results (as bitstreams, post-P\&R layout or mask layout files), but also, as much as possible, release the higher level source code. 

% \newpage
%%%%%%%%%%%%%%%%%%%%%%%%%%%%%%%%%%%%%%%%%%%%%%%%%%%%%%%%%%%%%
\section*{Acknowledgments}
The DRAC project is co-financed by the European Union Regional Development Fund within the framework of the ERDF Operational Program of Catalonia 2014-2020 with a grant of 50\% of total cost eligible. We also thank Red-RISCV for the efforts to promote activities around open hardware.
% Acknowledgements removed due to anonymous reviewing.

%% To balance last two columns
% \newpage

\bibliographystyle{IEEEtran}
%\bibliography{dcisbibliography}
\bibliography{conference_101719}

\end{document}